\documentclass[graybox, nosecnum]{svmult}

\usepackage{mathptmx}       
\usepackage{helvet}         
\usepackage{courier}        
\usepackage{type1cm}        
\usepackage{makeidx}         
\usepackage{graphicx}        
\usepackage{multicol}        
\usepackage[bottom]{footmisc}
\usepackage{hyperref}        
\usepackage{soul}            
\hypersetup{colorlinks=true,urlcolor=blue}
\usepackage{subcaption}
\usepackage[square,numbers]{natbib}
\makeindex             

\begin{document}
\title*{Optical Networks and Interconnects}
\author{Carmen Mas-Machuca \thanks{corresponding author}, Lena Wosinska, Marco Ruffini, and Jiajia Chen}
\institute{Carmen Mas-Machuca \at Technical University of Munich (TUM), Arcisstraße 21, 80333 München \email{cmas@tum.de}
\and Lena Wosinska \at Chalmers University of Technology, 41296 Gothenburg, Sweden \email{wosinska@chalmers.se}
\and Marco Ruffini \at CONNECT centre, School of Computer Science and Statistics, Trinity College Dublin, Ireland \email{Marco.Ruffini@tcd.ie}
\and Jiajia Chen \at Chalmers University of Technology, 41296 Gothenburg, Sweden \email{jiajiac@chalmers.se}}
%
%
\maketitle
\abstract{The rapid evolution of communication technologies such as 5G and beyond, rely on optical networks to support the challenging and ambitious requirements that include both capacity and reliability. This chapter begins by giving an overview of the evolution of optical access networks, focusing on Passive Optical Networks (PONs). The development of the different PON standards and requirements aiming at longer reach, higher client count and delivered bandwidth are presented. PON virtualization is also introduced as the flexibility enabler. Triggered by the increase of bandwidth supported by access and aggregation network segments, core networks have also evolved, as presented in the second part of the chapter. Scaling the physical infrastructure requires high investment and hence, operators are considering alternatives to optimize the use of the existing capacity. This chapter introduces different planning problems such as Routing and Spectrum Assignment problems, placement problems for regenerators and wavelength converters, and how to offer resilience to different failures. An overview of control and management is also provided. Moreover, motivated by the increasing importance of data storage and data processing, this chapter also addresses different aspects of optical data center interconnects. Data centers have become critical infrastructure to operate any service. They are also forced to take advantage of optical technology in order to keep up with the growing capacity demand and power consumption. This chapter gives an overview of different optical data center network architectures as well as some expected directions to improve the resource utilization and increase the network capacity.}
\section{Keywords} 
Access networks, core networks, optical datacenters, network virtualization, network planning, datacenter architectures, resource disaggregation, EON, LR-PON, energy efficiency.
\section{Introduction}
The strict quality of service requirements of emerging services and the rapidly increasing number of connected users and devices are bringing many challenges for both the wireless/radio segment and the wired segments interconnecting mobile base stations, which span from optical access to aggregation and core networks. This trend is expected to accelerate in future generations of mobile systems that will certainly pose extreme challenges to provide the required ultra-high transmission capacity, ultra-low latency and high reliability. Optical networks have long played a central role in telecommunication networks, forming the fiber backbone of the Internet. Over the years the penetration of fiber increased rapidly in order to support ultra-high capacity, ultra-low latency, and high reliability performance requirements of the services, provided by current and future generations of mobile networks. Thanks to the unique properties of optical transmission, photonic components and devices, as well the rapid development of optical network equipment, optical networks and interconnects have evolved into flexible network infrastructures able to support growing capacity demand and the challenging service requirements. There is no doubt that optical networks together with the next generations of mobile networks and the proliferation of artificial intelligence throughout the network, will play a key role in the future communication networks.

Moreover, all the modern services are in one way or another processed in data centers, leading to a dramatic growth of data center traffic. This brings serious challenges for the intra-data center interconnects in terms capacity demand and power consumption. In this respect, photonics and optical transmission are the only viable technologies for the intra-data center interconnects.

This chapter will guide the reader through different segments of optical networks and interconnects, providing an overview of their developments and future trends. The chapter is divided in four parts. The first section is dedicated to optical access networks, which are crucial for the development of new generations of mobile networks and services. The second part is focusing on optical core networks and addresses various network planning problems, as well as optical network control and management. The third part is concentrating on optical intra-data center interconnects, describing possible architectures, highlighting the main challenges and future developments. Finally, concluding remarks are given in the fourth part. 

\section{Optical Access Networks}
This part provides an overview on the architectural evolution of optical access networks, focusing on Passive Optical Networks (PONs). This section starts by describing the evolution in PON standards. Then it delves into architectural evolution, proposed in the research literature, to improve capacity, cost effectiveness and support for new services, namely ultra dense WDM, Long-Reach PON and mesh PON architectures (some time referred to as EAST-WEST or inter-Optical Network Units (ONUs)). It then describes the current trend of PON virtualisation, emphasising the increase in flexibility that this can provide.

\subsection{Overview of optical access networks architecture evolution}
\subsubsection{Evolution of classical PON architectures}
The concept of bringing fibre connectivity all the way to the home (i.e., Fibre to the Home - FTTH) has been around for several decades, with first trials appearing as far back as 1977 in Japan \cite{FTTH_77}. It was soon realised that scalability was going to be the main issue, as access technology needs to scale to tens or hundreds of million connections for each operator. Network infrastructure sharing was identified as a viable option for reducing cost, especially as it suited well the highly statistical nature of end user capacity demand and utilisation.
This led to the development of the Passive Optical Network (PON) concept, where infrastructure sharing is achieved through the use of passive power splitters, thus creating a broadcast and select type of architecture. The low loss and high capacity of optical fibre transmission was a key enabler for such passive split system.

Over the years, PON technology and standardisation have evolved, providing increasingly larger capacity, which was required to support increasing bandwidth demand from multi-media applications.
Starting from the Telephone PON (T-PON), offering the equivalent of 128 ISDN (144 Kb/s) connections from a single feeder fibre, ITU-T standards have evolved towards Broadband PON (B-PON), offering 155Mb/s upstream and 622 Mb/s downstream, Gigabit PON (G-PON) offering 1.25 Gb/s upstream and 2.5 Gb/s downstream and XG-PON (i.e., 10G-PON, 2.5 Gb/s upstream and 10 Gb/s downstream). The rates provided here in brackets were the most widely used, as standards typically allow a range of possible transmission rates. In addition, all such technologies are based on Time Division Multiplexing (TDM), and often referred to as TDM-PON systems.
It should be noticed that some of the standards, in particular (G-PON and XG-PON) have offered asymmetric upstream/downstream rates. Unlike copper-based Asymmetric Digital Subscriber Line (ADSL), where such asymmetry was due to the limited bandwidth of the medium, for PONs this was due to the higher complexity of the Optical Line Terminal (OLT) receiver, which needs to operate in burst mode in order to enable upstream multiplexing of burst data transmission from the end user Optical Network Units (ONUs).

After XG-PON, standardisation moved towards multi-wavelength solutions, with NG-PON2 (Next-Generation PON 2), offering four Dense Wavelength Division Multiplexing (DWDM) channels with symmetric 10 Gb/s capacity, hence the name of Time-Wavelength Division Multiplexing PON (TWDM-PON). In practice though, NG-PON2 saw little deployment due to the high cost of the ONUs (which required tunable laser and filters). For this reason a new standard XGS-PON (10G Symmetric PON) was developed to bring a symmetric 10Gb/s PON, with an upgrade path towards the multi-wavelength NG-PON2 system. 
 In the meantime the evolution towards higher data rate channels has progressed, with ITU-T HS-PON (High-Speed PON) working group, providing 25 Gb/s upstream and 25 to 50 Gb/s (still under investigation) downstream.

Another standardisation body, the IEEE, has also defined PON standards (the E-PON - Ethernet PON family), built upon the Ethernet standard. IEEE and ITU-T PON standards are similar in architectures and data rate evolution, with the ITU-T providing additional quality of service management features in the upstream.

In parallel with these standardisation efforts, other architectures have been proposed and implemented through proprietary solutions, such as Wavelength Division Multiplexing (WDM) PON. The main difference with respect to TDM PON is the use of a wavelength multiplexer instead of the power splitter (although mixed use of multiplexer and splitter is also found in some architectures). The idea is to provide one dedicated wavelength channel to each end user: thus fibre infrastructure is shared through WDM, but each user has a dedicated OLT termination. This concept was further developed in proposals such as Ultra-Dense WDM PON (UD-WDM-PON) \cite{UDWDM}, where coherent technology was developed to provide in the order of a thousand different channels. The main reason hindering its development was similar to that for NG-PON2, i.e., the high cost of tunable transceivers, in addition to requiring one OLT device for each end user. While this can potentially offer high-capacity dedicated channels to end users, it should be noticed that computer networks are always built upon statistical multiplexing principles, in order to remain financially viable and scalable. Thus, while WDM-PON can provide large uncontended capacity to each end user, such capacity is multiplexed at the central office (CO), where contention inevitably occurs. The advantage of TDM (or TWDM)-PONs is that statistical multiplexing is carried out closer to the end users, improving resource efficiency. At the same time, the broadcast nature of the power split architecture enables the overlay of dedicated wavelength connections for specific users, for example if a business requires a dedicated 25G connection it will bear the extra cost for a more complex ONU and a dedicated OLT.

\
\subsubsection{Long-Reach PON architectures}
An interesting architecture that has been considered for long time (it was initially proposed in 1990 \cite{Hill_90}), is the Long-Reach PON (LR-PON) \cite{LR-PON}. It's distinctive feature is that it adopts a much longer optical reach and split ratio (i.e., number of ONUs) than conventional PON architectures. The main rationale behind the LR-PON concept is that today's network topology, with COs often within a few km of any end user, was driven by the reach limitation of the copper transmission technology when cables were laid out over a hundred years ago. The concept behind LR-PON is to avoid using fibre just as a mere replacement for the transmission medium, and instead use its extraordinary transmission capability to provide a complete redesign of the network architecture, thus delivering more cost effective increase of overall network capacity. The key idea is to exploit the low transmission loss, high bandwidth and low cost amplification of optical technology to design PONs able to cover transmission distances above 100 km and split ratios up to 1024. 

\begin{figure}
	\centering
	\begin{subfigure}{.45\textwidth}
 	 \centering
	  \includegraphics[width= 0.9\linewidth]{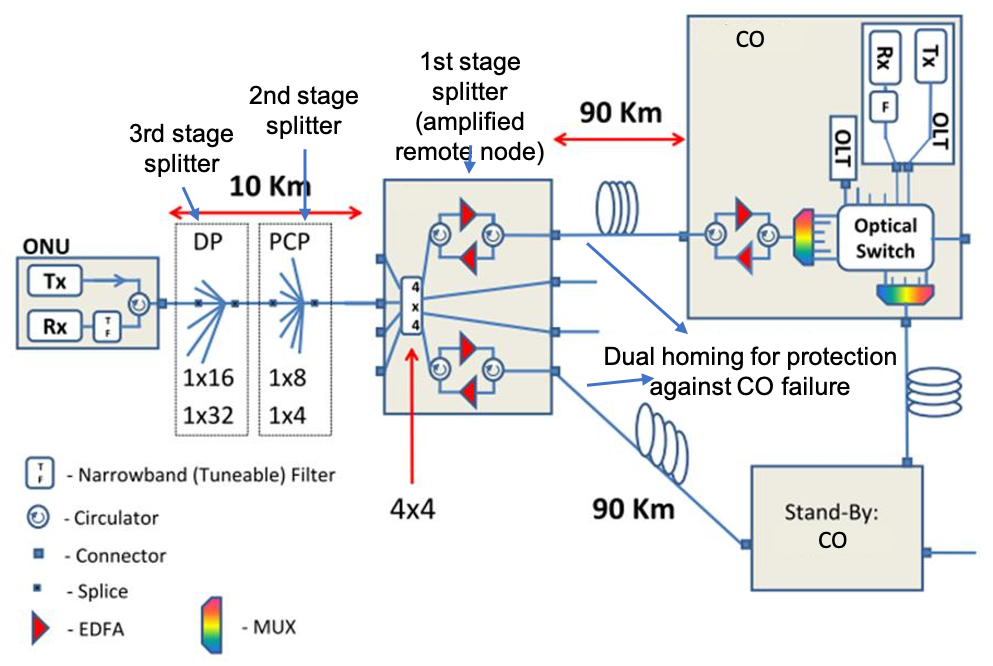}
	\end{subfigure}%
	\begin{subfigure}{.55\textwidth}
	  \centering
 	 \includegraphics[width= \linewidth]{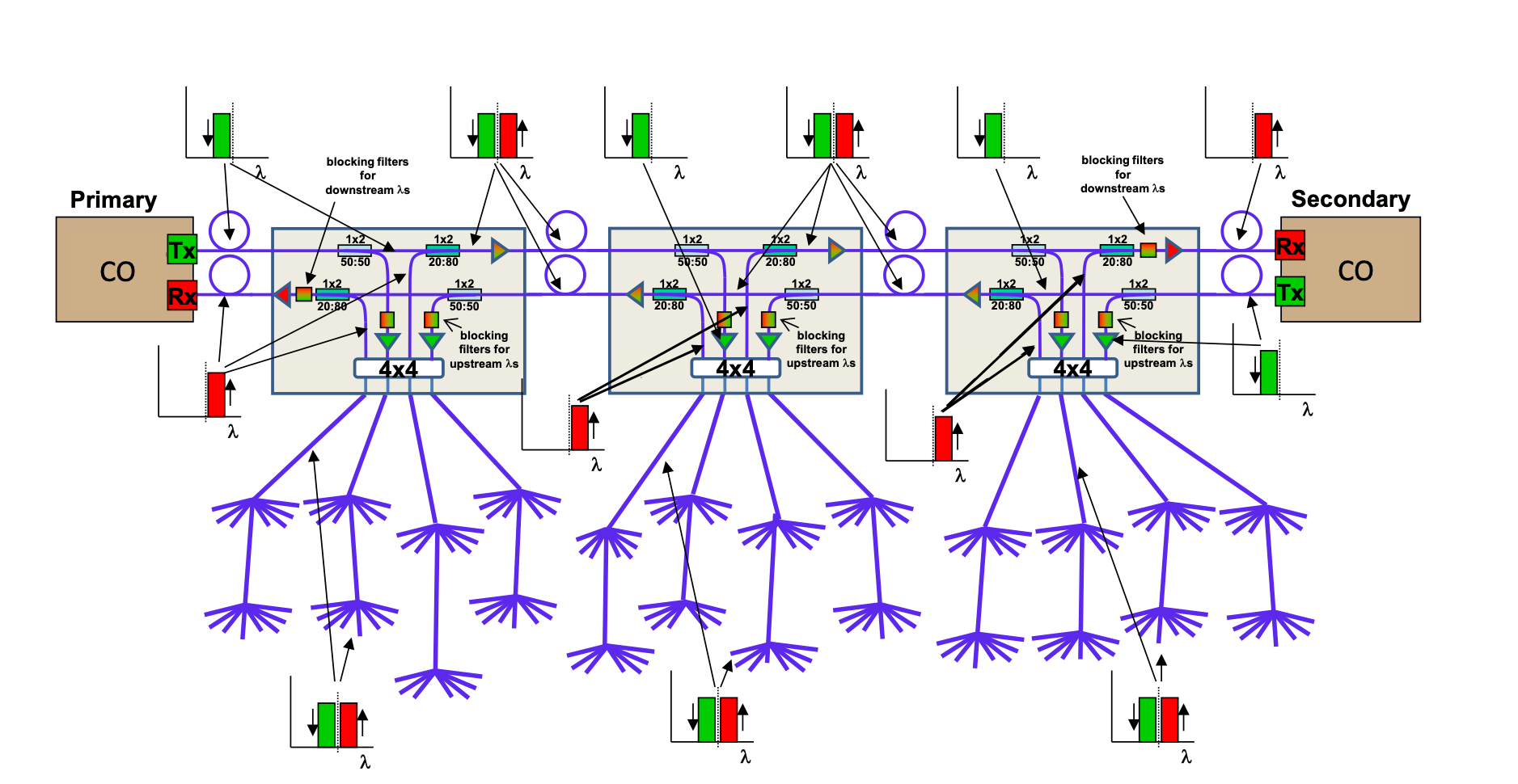}
	\end{subfigure}
	\caption{Two different LR-PON architectures. On the left hand side, a tree-based topology with long feeder fibre and 10 km ODN; on the right hand side a ring-based topology where splitter nodes are distributed across a wide area.}
	\label{fig:LRPON_arch}
\end{figure}

This architecture brings about two main changes in the network. It enables central office consolidation, reducing their number, typically by almost two orders of magnitude. It simplifies the core network, as the reduction in number of COs brings about the possibility to provide a flat, single tier core network (i.e., without electronic hops) connecting all COs through a mesh topology of wavelength channels \cite{Flat-core}. 
Two main LR-PON architectures are shown in Figure~\ref{fig:LRPON_arch}. One providing a long feeder fibre (90 km), followed by a short (10 km) Optical Distribution Network (ODN). This is shown on the left hand side of Figure~\ref{fig:LRPON_arch}, where the first stage splitter provides optical amplification and two additional splitter locations further distribute the signal, typically following existing duct routes. Another architecture, more suitable for lower density scenarios (i.e., rural areas) is the ring feeder, shown on the righ hand side of Figure~\ref{fig:LRPON_arch}, where multiple COs are connected through a feeder ring.

\begin{figure}
 	 \centering
	  \includegraphics[width= \linewidth]{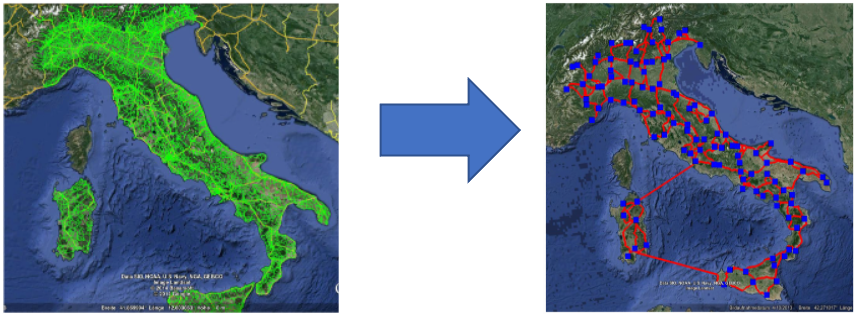}
 	 \caption{LR-PON central office consolidation for the Italian network use case.}
	  \label{fig:consolidation}
\end{figure}

As mentioned above, this enables a simpler core network, where the only electronic packet switching is carried out at the COs, with no need for intermediate electronic packet routing within the network domain.
Figure~\ref{fig:consolidation} shows how the number of COs could be reduced by two orders of magnitude, for a scenario built around Italy's national network, moving form over 10,000 COs, down to 116 nodes.

 The diversity of this architecture also enables novel approaches to its control and operation. For example, the larger geographic span of each CO (now covering entire metropolitan areas) can facilitate higher infrastructure sharing for protection purposes, as proposed in \cite{Prot1} and demonstrated in \cite{Prot2}.
  
  A key element for the cost-effectiveness of the LR-PON architecture shown in Figure \ref{fig:LRPON_arch}, which enables the long-reach feature, is the amplification at remote nodes (i.e., the first stage splitter). Here an optical amplifier (typically an Erbium Doped Fibre Amplifier- EDFA, but Semiconductor Optical Amplifiers - SOAs can also be adopted, as shown in \cite{LRPON-SOA}). Just two amplifiers (one for each direction) can provide enough gain for multiple wavelength channels, (40 bidirectional channels were demonstrated in \cite{LRPON-demo} operating in the C band). Taking in consideration the tree-based architecture in Figure \ref{fig:LRPON_arch}, for example, the power budget can be of the order of ~25 dB for the feeder, which is compensated by the amplifier, before entering the ODN section, where a split of 1024 and a reach of 10 km are supported with a power budget of ~30 dB.
  
One of the main drawbacks of this architecture, which has hindered its practical development, is that it provides a level of centralisation that is in antithesis to the latest trend of edge cloud computing, which requires active networking and computing nodes close to the end user. This is driven by both 5G application requirements (i.e., ultra-low latency, which often require use of local edge nodes for processing) and 5G networking requirements, where low latency is required to support cloud Radio Access Networks (C-RAN), functional split architectures \cite{NGMN}.

\subsubsection{PON evolution in support of small cells and edge computing}
One of the major approaches of 5G to capacity increase is Radio Access Network (RAN) densification, through dense deployment of small cells. While macro cells and other computing or service nodes have been typically connected by point-to-point fibre links, as 5G increases their number by orders of magnitude, point-to-point fibre solutions become too expensive. Recently, PONs have been considered as a more viable option to connect 5G small cells and Multi-Access Edge Computing (MEC) nodes \cite{PON_backhaul}. 
A key advantage of PONs is that they are cost effective by means of providing statistical multiplexing. As small cells cover smaller areas and thus a smaller number of users at a time compared to larger macro cells, their load will fluctuate more over time, making them a good match for TDM (or TWDM) PONs, which provide statistically multiplexing ability.
However, since PONs were initially designed to offer broadband connectivity to residential users and small businesses, some work is needed to improve their performance to match the 5G network requirements, in terms of latency, jitter and capacity.

\begin{figure}
 	 \centering
	  \includegraphics[width= \linewidth]{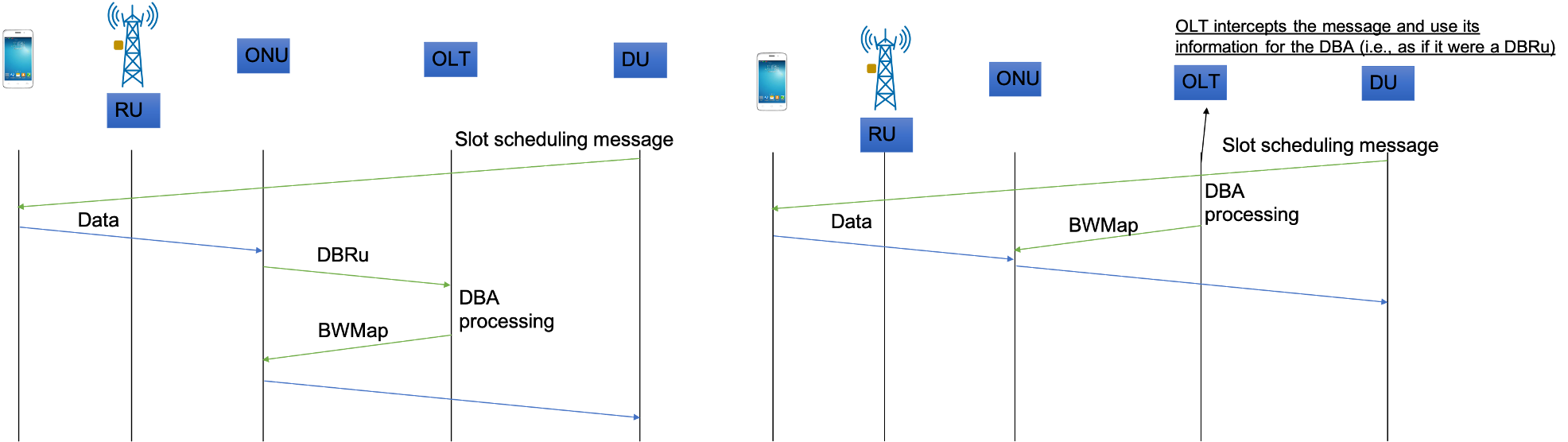}
 	 \caption{Standard DBA operation for a PON fronthaul application (left) and Cooperative DBA solution for coordinating fixed and mobile scheduler (right).}
	  \label{fig:Co-DBA}
\end{figure}

From a capacity perspective, as mentioned above, current standards will soon be delivering 50Gb/s, while higher rates of 100 Gb/s and beyond are being investigated. In addition, multi-wavelength systems are envisaged to be a necessity if PONs are to provide fronthaul RAN transport services. While multi-wavelength systems proved to be too expensive for residential services, their cost can be justified for 5G small cell, especially considering the fact that the alternative of leasing point to point fibre can be much more expensive.
Another issue is that current PONs cannot easily deliver latency below 1 ms. When considering fronthaul transport, for example the popular OpenRAN 7.2 split, latency requirements are of the order of 100 $\mu s$ between the Remote Unit (RU), where the antenna is located, and the Distributed Unit (DU), where the signal baseband processing is located \cite{ORAN-latency}. A solution to this problem was initially proposed in \cite{NTT-OFC-2014}, called Cooperative DBA, where it was recognised that the fronthaul latency issue could be eliminated if both RAN and PON schedulers could be synchronised. The difference in operation compared to a standard PON is shown in Figure~\ref{fig:Co-DBA}. The left hand of the figure shows a typical PON DBA operation, where the ONU needs to transmit a report indicating its upstream queue level to the OLT, so that capacity can be scheduled for the next cycle. This operation is time consuming, so if the ONU serves a small cell, the incoming packet needs to wait in the queue until this process is completed and a transmission opportunity is granted. In Cooperative DBA, shown on the right hand side of the figure, the small cell scheduler provides advanced information on when the end user (i.e., the handset) is scheduled for upstream transmission, so that the time such data will arrive at the ONU can be accurately estimated. This enables the OLT to calculate the bandwidth map in advance and thus provide the ONU with an upstream transmission opportunity a few $\mu s$ after the arrival of the packet from the small cell (which is directly connected to the ONU). This concept was recently standardised at the ITU-T as Cooperative Transport Interface (CTI) \cite{CTI}. \par

\begin{figure}
 	 \centering
	  \includegraphics[width= 0.8 \linewidth]{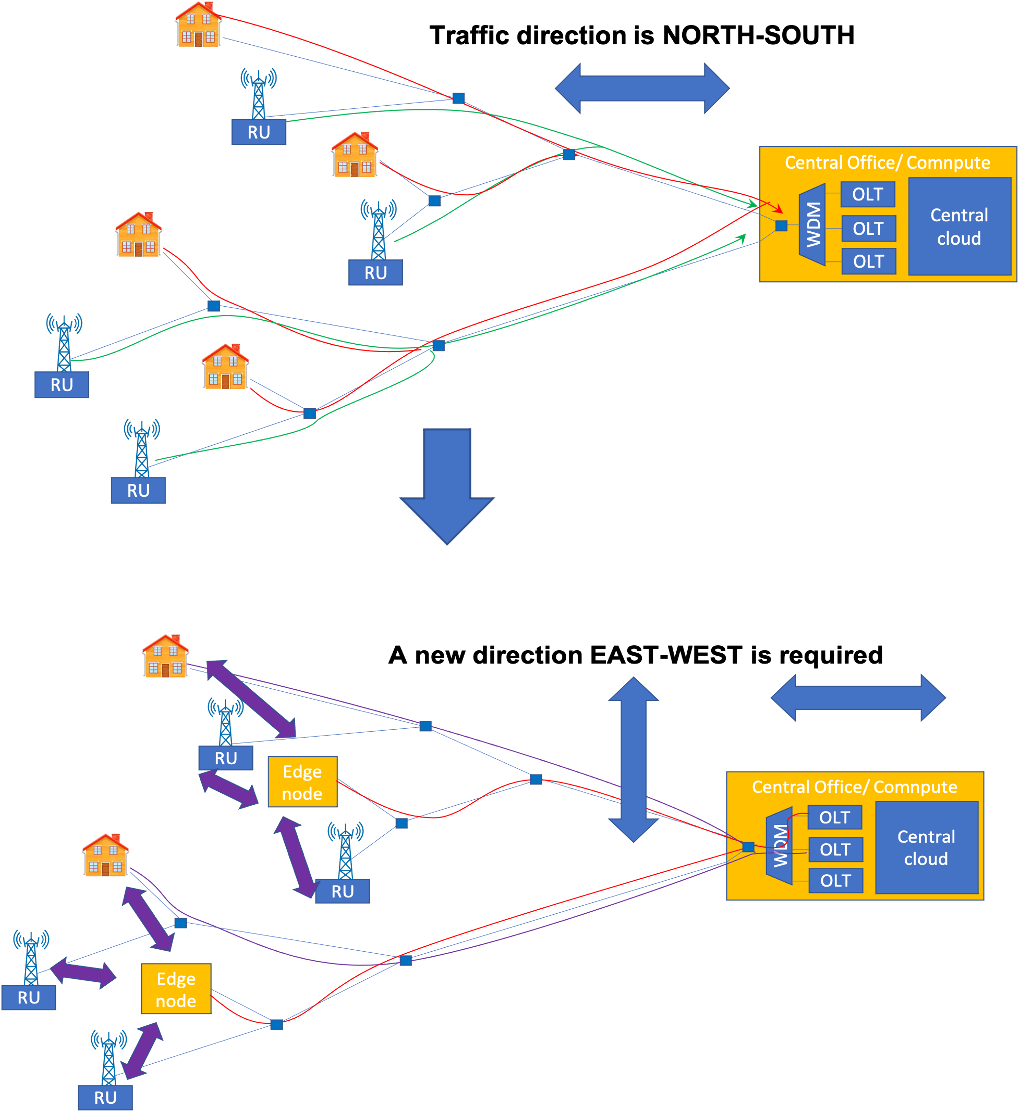}
 	 \caption{The introduction of MEC nodes in the network changes the traffic patterns into a mesh demand matrix.}
	  \label{fig:mesh_traffic}
\end{figure}

Another aspect of 5G evolution is that of edge computing: the high capacity and low latency demand posed by 5G network services and applications has driven the development of the edge cloud, i.e., moving computing capacity closer to the end user. There is a multitude of envisaged scenarios, for example: computation offload from mobile handset (and smart glasses in the near future) for applications such as real time object recognition; support for autonomous driving and smart traffic intersections; support of cloud-RAN services in high densification scenarios, etc.
While edge computing nodes can provide such localised computing facilities, the issue remains on how to interconnect them with the rest of the access network. Due to their envisaged high density (which could be similar to the number of small cells in a 5G access network), and their mesh traffic patterns (i.e., MEC nodes will need to communicate with multiple 5G cells, towards COs and to other MEC nodes, as shown in Figure~\ref{fig:mesh_traffic}), providing cost effective connectivity remains a major challenge.

If the MEC architecture has a hierarchical structure that can be mapped into a tree topology, for example a given MEC serving a number of small cells, then overlapping PONs could be considered. For example, a first-stage PON connects the CO to a number of splitters, some of which could host MEC nodes, which, in turn, could operate a second-stage PON network connecting to the small cells. However this solution is static and prescriptive, i.e., it does not allow an MEC node to support other small cells (i.e., for load balancing), or to other MEC nodes (i.e., for sharing information at lower latency, with respect of being routed through a CO).

\begin{figure}
 	 \centering
	  \includegraphics[width= \linewidth]{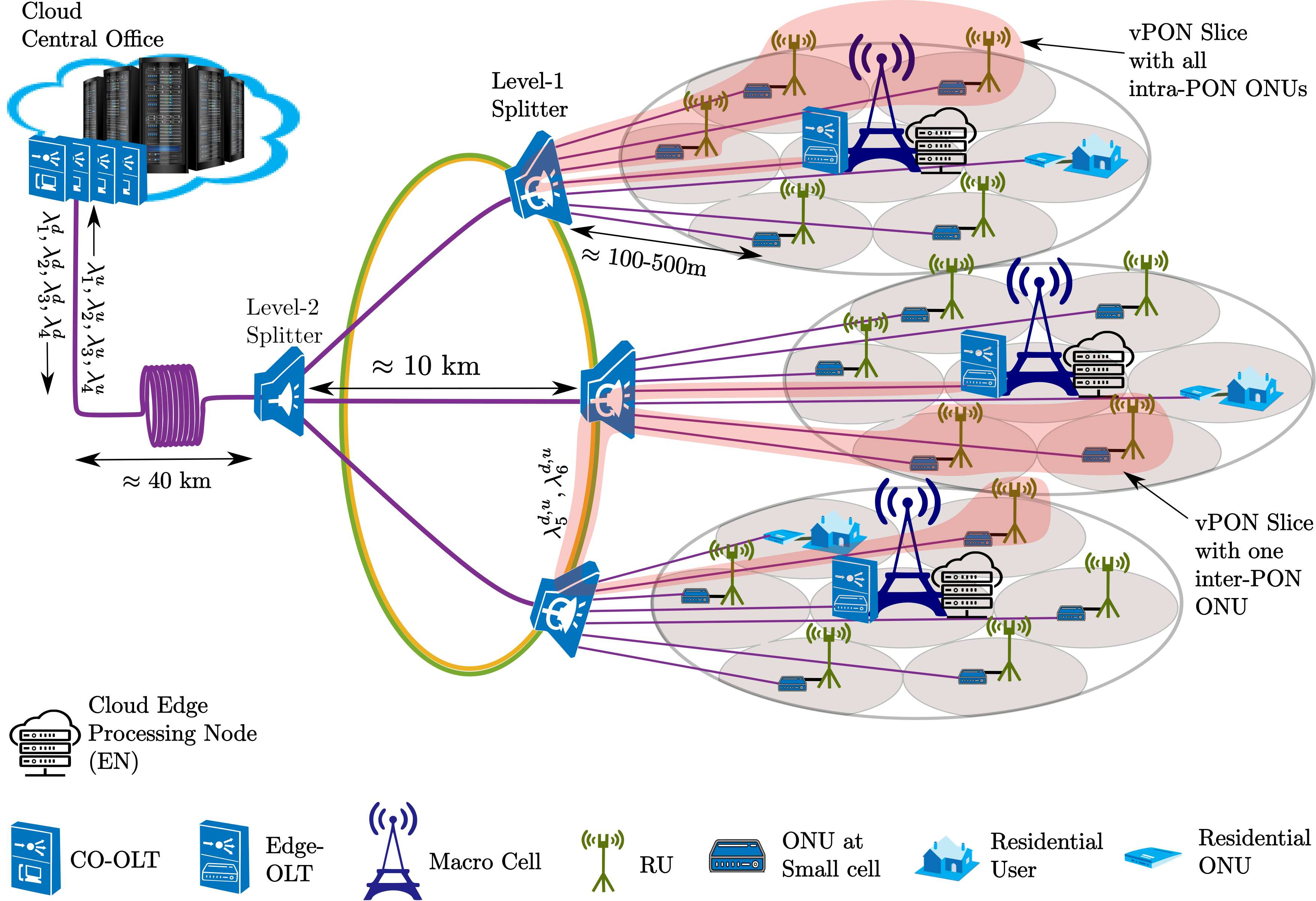}
 	 \caption{Mesh-PON architecture enabling direct communicaiton across end points, through wavelength reflection at splitter location.}
	  \label{fig:mesh_PON}
\end{figure}

Other solutions are based on more flexible architectures that enable direct communications across end points \cite{Intra-ONU1}, \cite{Intra-ONU2}. 
Figure~\ref{fig:mesh_PON}, shows a solution based on the use of tunable Fibre Bragg Gratings \cite{MESH-PON}, to selectively reflect wavelength channels towards the end points, thus enabling direct communication between them. The idea is that an OLT can be located at an end point and one of its wavelengths be reflected back by one of the splitters towards the other ONUs, creating a local communication group. By selecting which wavelength to let through and which to reflect, at each splitter, multiple such local communications groups can be created, even across different PON branches \cite{MESH-experiment}. This type of architecture effectively adds the EAST-WEST direction of communication to a PON, in addition to the classical NORTH-SOUTH, thus enabling mesh connectivity between MEC nodes and towards small cells or other end points.
One of the issues of this model however, is that as the signal traverses the splitters more than once, they get attenuated. Thus, connecting end points across multiple branches will likely require optical signal amplification (i.e. EDFAs), as they will traverse multiple splitters. 

\subsection{Optical access network virtualisation}

The emergence of 5G was accompanied by another major revolution in networking, the virtualisation of network functions, known as NFV (Network Function Virtualisation).
While this is a vast topic that involves a large pool of technologies, from data centres to telecommunications networks, this section briefly discusses how virtualisation has affected the optical access network domain \cite{Cloud-access}. 
While the use of virtualisation was applied to PCs and servers for a long time, the concept has reached the networking realm relatively recently. This occurred through NFV, where network functionalities typically implemented in proprietary equipment, have moved to run as software instances over general purpose servers. In telecommunications networks, the key milestone occurred when virtualisation reached the Central Office. The first project to go towards this direction was the Central Office Rearchitected as a Data Centre (CORD) \cite{CORD}, which aimed at running services in a Telecomms central office as open source software on general purpose servers and whitebox switches and routers (i.e., open systems based on merchant silicon). 

There are several reasons behind this move towards virtualised systems. Some of the main advantages are:
\begin{itemize}
\item encourage a competitive environment, which typically leads to cost reduction.
\item increase interoperability among systems, leading to increase in functionalities and avoidance of vendor lock in.
\item provide a higher degree of programmability, leading to increased customisation.
\item enable slicing, leading to better support for multi-tenant and multi-service environments.
\end{itemize}

The CORD architecture was initially partitioned into three use cases: Residential (R-CORD), Mobile (M-CORD) and Enterprise (E-CORD). Over time these have evolved and delivered new projects. For example, the SDN enabled broadband access (SEBA) is a lightweight variant of R-CORD, supporting fixed broadband technologies such as PON, G.Fast (delivering fibre to the cabinet) and DOCSIS (broadband over coaxial cable). The Converged Multi-Access and Core (COMAC) project was developed to integrate mobile and fixed access with the network core. AETHER is a new platform extending the CORD concept to support edge cloud as well as private 5G connectivity.

This section focuses on the virtualisation aspects related to PONs. While the SEBA project provides the framework required to run a fixed access network, the Virtual OLT Hardware Abstraction (VOLTHA), provides the control and management aspect of a PON access system. VOLTHA's northbound interfaces allows the PON OLT to be abstracted as an Ethernet switch to the network controller (for example the ONOS controller developed by the Open Networking Foundation -ONS). The southbound interface instead connects directly to the PON whitebox device.
This virtualisation enables the creation of PON slices, where different tenants can share the same PON infrastructure with the total PON capacity shared across multiple PON slices. The system has been successfully demonstrated for operation in typical residential PON use cases, including production deployment by the German operator DT \cite{DT-SEBA}.

 \subsubsection{Deep PON virtualisation}
While the abstraction level provided by VOLTHA and SEBA addresses the current needs of residential PON customers well, 5G is bringing about new requirements. 
As mentioned above, PONs are being considered as a solution to reduce the cost of mobile small cell densification, which uses functional split architectures. This makes use of lightweight remote antenna devices called Remote Unit (RU), connected via fibre to edge compute nodes or central offices, where the rest of the digital signal processing is carried out in Distributed Units (DU) and Centralised Units (CU) \cite{NGMN}. While this architecture provides the benefit of computing centralisation, as only a minimum amount of processing is carried out at the RU site, it poses strict requirements to the interface between the RU and DU, in particular, it requires a latency of the order of 100 $\mu s$. Since the RU is located at the very edge of the network, PON connectivity, where already deployed for other residential uses, can provide high capacity and reliable connectivity at low cost. However, the latency requirement of the RU-DU interface is not supported by standard PONs, as the upstream scheduling mechanism typically operates on the timescale of a few ms. As mentioned above, the Cooperative Transport Interface provides a solution to this problem, and virtualisation of the PON scheduler, which goes beyond the VOLTHA implementation, can facilitate its implementation. An example of PON scheduler implementation was proposed in the FASA architecture \cite{FASA}, standardised in \cite{BBF-402}.
However, operating in a multi-tenant and multi-service environment, i.e., sharing a physical PON across multiple mobile and residential services, requires each tenant to control a scheduler instance.

\begin{figure}
 	 \centering
	  \includegraphics[width= 0.7 \linewidth]{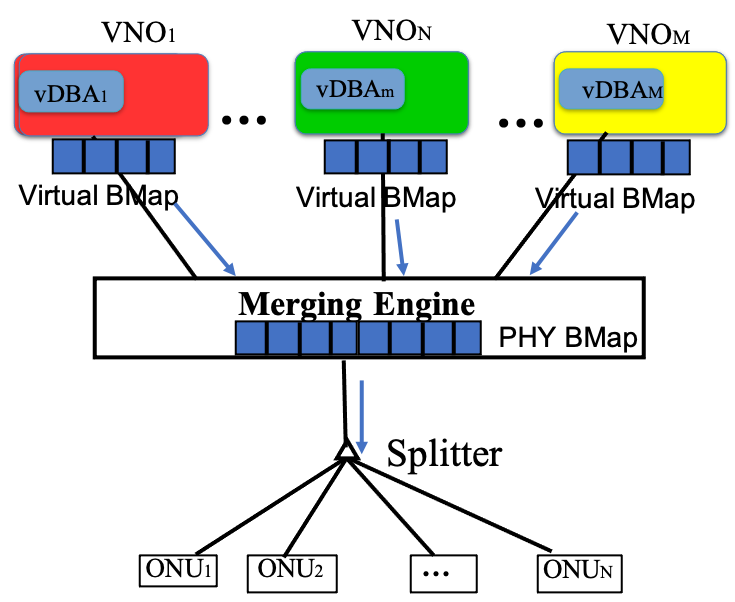}
 	 \caption{Virtual DBA mechanism for multi service and multi tenant operation. Each Virtual network operator (VNO) can calculate an independent bandwidth map allocation, which then is merged into a common physical bandwidth map that is sent to all ONUs.}
	  \label{fig:vDBA}
\end{figure}

This concept was developed in \cite{vDBA-concept}, and standardised in \cite{BBF-402}, \cite{BBF-370}, with the name of virtual Dynamic Bandwidth Allocation (vDBA), where each different tenant (or service) can operate an independent upstream scheduler (DBA) allocating a capacity up to a given maximum level. 
Following the example in  Figure~\ref{fig:vDBA}, each vDBA would calculate an independent upstream scheduling allocation (called Bandwidth Map - BWMap) and send it to a merging engine (ME), which has the task of merging all upcoming BWMaps, while resolving any collisions in the allocations.

The merging engine thus becomes the critical element for supporting QoS.
Initial approaches to the ME have followed relative prioritisation, typical of standard stateless Ethernet/IP queue management approaches, where packets from services with higher priority are given preference over packets from lower priority services \cite{vDBA-test}. The issue with simple packet prioritisation is that it is based on relative importance between flows. However, this does not operate over the key performance metrics of a service level agreements (SLA), such as latency and jitter, thus it acts blindly and can be difficult to control, especially with respect to flows that have similar level of priority. 
A solution was proposed in \cite{OFC-21}, where a ME was designed to operate directly on SLA parameters for next-generation services, for example that a given latency of 100 $\mu s$ is maintained for 99\% of upstream slot allocations, measured over a given time window. An example of a second SLA-oriented service could require a latency of 50 $\mu s$, but with 95\% availability. If relative prioritisation was to be used, it would be difficult to determine which of the two services should be given higher relative priority, so the key approach here is to act on the actual SLA parameters.
The solution proposed in \cite{OFC-21} requires the use of a stateful mechanism, meaning that the algorithm that implements the merging process keeps track of the current and past performance (the state) of all services that require QoS (i.e., SLA-aware). Any time there is a scheduling collision between two DBA allocations, the algorithm checks the flow delivery statistics and gives priority to the one that is closer to breaching its SLA. The effect of this algorithm is to minimise the chance for any service to break its SLA, which is the main concern that a PON operator would have in providing SLA-oriented 5G services.
In this way, an operator could carry out simulation analysis for different traffic profiles and SLA types to estimate the maximum number and types of services the PON can support before leading to SLA breach.


\section{Optical Core Networks}
Telecommunication networks are commonly categorized into three segments: core, aggregation and access networks. Core networks are the most capacity hungry segment as its demands are between Internet Exchange Points (IXPs), data centers and big cities. The most common topology of core networks is bi-connected mesh topology, which allows interconnecting any two nodes with at least two disjoint paths. This requirement allows operators to increase the robustness of their network in case of single failures. Indeed, this requirement is associated with the high capacity of optical links. 

The first optical core networks were opaque, that is, the signal transmission was optical, whereas the switching was performed in the electrical domain. Converting the signal to the electrical domain at each node required a large investment due to the high number of transceivers. In order to reduce the cost, translucent or even transparent optical core networks have been implemented such that not only the transmission but also the switching is performed in the optical domain. Transparent optical networks not only reduce the number of required transceivers, but also the consumed power and the control data to configure the lightpath. 
\begin{figure}[h]
\centering
\includegraphics[width=0.85\textwidth]{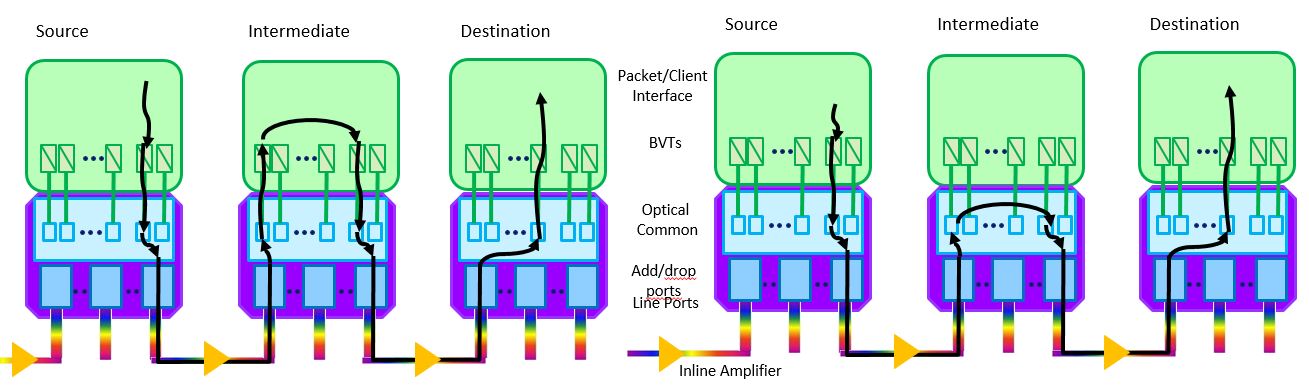}
\caption{Opaque (left) vs. Transparent (right) optical networks~\cite{Achim2011}}
\label{fig:opaque}
\end{figure}

Optical links consist of optical cables with tens or hundreds of fibers. The capacity of each fiber depends on the multiplexing technique. In core networks, Wavelength Division Multiplexing (WDM) is used to increase the capacity according to the number of wavelengths (e.g., Dense WDM (DWDM) supports 40 channels at 100 GHz spacing or 80 channels with 50 GHz spacing, whereas Ultra Dense WDM (UDWDM) supports hundreds of channels with 12,5 GHz spacing~\cite{Li2018}). As the link distances in core networks are large (hundreds of kilometers), amplification is required, resulting in heterogeneous span lengths from 40 to 120 km ~\cite{Chin2017}. Furthermore, the node components have also evolved: from opaque nodes performing the switching in the electrical domain, towards transparent optical Reconfigurable Optical Add/Drop Multiplexers (ROADMs). Current ROADMs have the following properties: Colorless (i.e., any color can be added/dropped at any port), Directionless (i.e., any color can be sent to any fiber/direction) and Contentionless (i.e., any color can be added/dropped in different directions simultaneously). 

Triggered by the high bitrate of the demands in core networks, Bandwidth Variable Transponders (BVT) allow setting lightpahts with a range of bitrates from 100 to 600 Gb/s~\cite{Sai2020}. Based on the lightpath characteristics (length, bitrate, etc.) the BVT will have to be configured accordingly in terms of modulation format, required spectrum, etc. 

Network operators are currently using the C-Band spectrum for their demands. However, when using the ITU fixed-grid, the number of lightpahts that can be set and the spectrum utilization are limited. In order to overcome these limitations, flex-grid combined with the use of BVTs, allow channels to use the spectrum they need according to their requirements (e.g., bitrate) while reducing wasted spectrum.  
In the next section will address the related planning problems more in detail. 

\begin{figure}[h]
\centering
\includegraphics[width=0.85\textwidth]{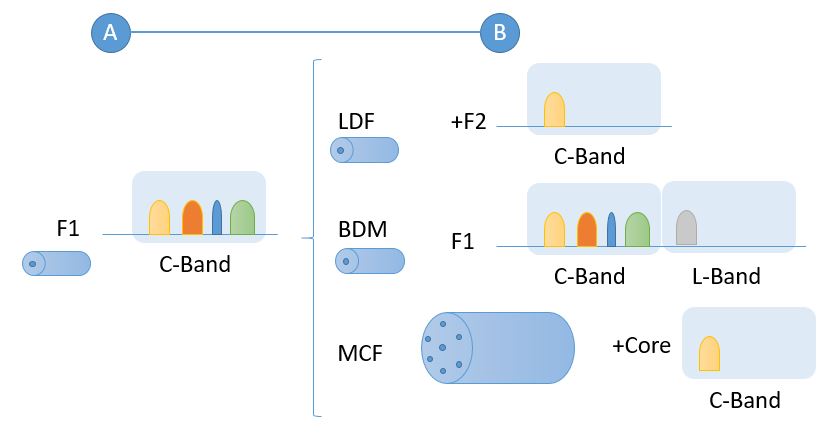}
\caption{Different alternatives to increase the capacity of a link: LDF, BDM, MCF}
\label{fig:more}
\end{figure}

Core operators are considering other possibilities to increase the capacity of their networks (refer to Figure~\ref{fig:more}: i) Lighting dark fibers (LDF) which uses the available dark fiber by lighting them with BVTs as needed; (ii) Band Division Multiplexing (BDM) uses other bands (e.g., C+L) in order to have broader spectrum although some equipment may be replaced or added to be compatible with the new bands; (iii) Multicore Fibers (MCF) replace the capacity of several SMF with one MCF. 

\subsection{Network planning problems}
\subsubsection{Routing and Spectrum Assignment problems} 
\begin{figure}[h]
\centering
\includegraphics[width=0.9\textwidth]{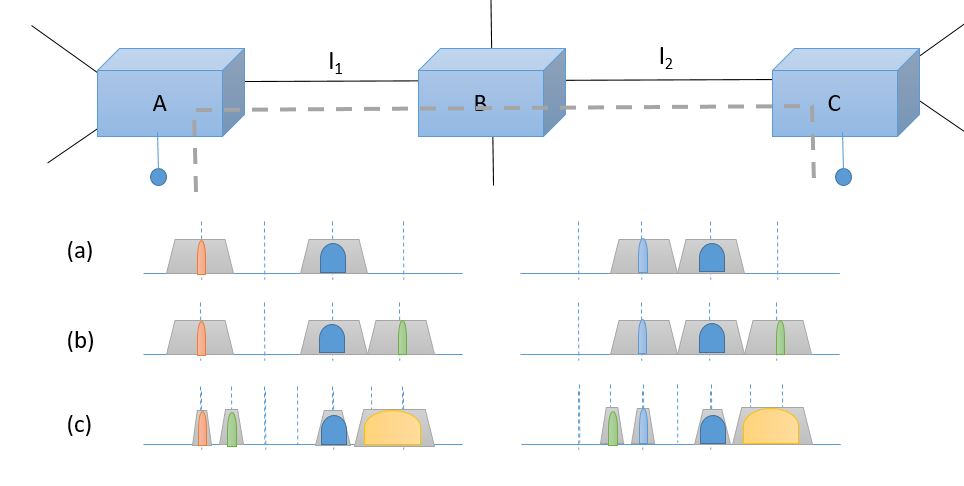}
\caption{Example of setting a demand from $A$ to $C$ through $b$ given the available spectrum shown in (a), when using (b) RWA and (c) RCSA}
\label{fig:RWA}
\end{figure}

The Routing and Wavelength Assignment (RWA) problem of WDM based networks without wavelength conversion, aims at finding the appropriate path and wavelength for a particular demand (defined by a required bitrate from a source node to a destination node). This problem has two constraints: i) the wavelength continuity constraint, which requires that the same wavelength can be used along all the links of the path; and ii) the wavelength assigned to one demand cannot be assigned to a second demand, which shares at least one link. These concepts will be shown with the example depicted in Figure~\ref{fig:RWA}(a) where three nodes ($A, B$ and $C$) are interconnected through links $l_1$ and $l_2$. For simplicity, let's consider 4 fixed grid slots of each link, two of them used for other demands with different bitrates. If a demand has to be set from $A$ to $C$ through $B$, the demand can be assigned to the last slot so that the two constraints can be guaranteed (as depicted in green in Figure~\ref{fig:RWA}(b)).  

With the development of flexible and configurable transponders, and the possibility to use the spectrum more flexibly and according to the demand's requirements, lightpath planning has evolved towards the Routing, Configuration, and Spectrum Assignment (RCSA) problem. 
The development of transponders supports software tunable channel configurations. Each configuration is associated to a particular combination of data rate, modulation format and FEC. Given a demand and a path, each configuration corresponds to a particular number of frequency slots to guarantee the minimum Optical Signal to Noise Ratio (OSNR) at the receiver. 
In elastic optical networks, where the spectrum is divided into smaller slots, the spectrum required by each demand should be mapped into a certain number of slots for all the links along the path. 
The RCSA problem aims at finding the appropriate path, configuration and spectrum for a particular demand. The problem has three constraints: i) the spectrum continuity constraint, which requires that the same spectrum slots are used along all the links of the path; ii) the spectrum slots assigned to one demand cannot be assigned to a second demand, which shares at least one link, and iii) the spectrum contiguity constraint, which requires that the spectrum slots are contiguous ~\cite{Sai2020, Amir2021}. Let's consider the example shown in Figure~\ref{fig:RWA}(a) where the low bit-rate demand from $A$ to $C$ should be set given the available spectrum. As the spectrum used for the demands is narrowed down to the required one based on the configuration, the new demand could be allocated between orange and light blue demands (shown in green in Figure~\ref{fig:RWA}(c)), leaving more contiguous available spectrum for future demands (e.g., the yellow demand with high bit rate as shown in Figure~\ref{fig:RWA}(c), which was not possible to be allocated in the Fixed Grid example of Figure~\ref{fig:RWA}(b)) . 

\subsubsection{Resilience}
Resilience is one of the most important requirements of optical core networks. 
As mentioned earlier, core networks have their nodes at operator locations, that are continuously monitored (e.g., temperature, humidity) and maintained.  However, links are installed along roads, train rails, etc. and are more likely to fail. Link failure rates range between 0.1 to 2.1 failures per 1,000 km per year~\cite{OFReport}, which means that a core network such as the European COST266~\cite{SNDlib} with 42,436 km, is expected to have tens of link failures per year.
Those links support demands of hundreds of Gb/s, which means that if the link fails (considering the number of fibers and the number of channels per fiber when using xWDM techniques), the interrupted traffic will be of the order of Pb/s. 

Fault management of optical networks consists of the following processes: 
\begin{itemize}
\item Failure detection: The performance monitoring of optical networks relies on the data monitored at the transceivers, as well as the monitored information from any controlled component and any signal monitoring (e.g., based on signal tapping). ~\cite{Kilper2004, Mrozek2018}. The performance monitoring of elastic optical networks is more challenging due to the complexity of the new components and the sensitivity to the physical impairments in order to efficiently use all the spectrum. The large amount of monitored data (e.g., one transceiver may have 10 monitored parameters every 30 seconds~\cite{Isabella2021}) enhances the problem complexity. Machine learning techniques are suitable to cope with such large amount of data and detect anomalies in a fast and accurate manner~\cite{Shahkarami2018, Faisal2018}. The accuracy is measured based on the false positive and false negative rates. 
\item Failure identification/localization: Once an anomaly/failure is detected, the exact identification of the type of failure and its location should be determined. For that purpose, different techniques such as machine learning~\cite{Panayiotou2018, Isabella2021} or heuristics~\cite{Vela2018} have been proposed.
\item Failure reparation: Once the failure has been identified, the reparation process is triggered, which depends on many parameters such as priority, available technicians and spare components, failure location, etc. Nevertheless, the operator is also concerned about the number and category of the interrupted demands. The SLAs signed with the clients, include a maximum demand unavailability~\cite{SLA}. When this value is exceeded, the operator should pay penalties which increase with the demand interruption time. Furthermore, the fierce competition encourages operators to comply with the client's requirements and keep their clients as long as possible. 
\end{itemize}

The objective of operators is to use their network resources efficiently while coping with the demand's requirements. Given a demand with a minimum availability, the operator should select the resources that will be assigned to that demand. There are two main categories of resilience schemes: protection and restoration. The former reserves some backup resources for a demand before any failure occurs. These backup resources may be dedicated to a particular demand (higher availability but high resource redundancy), or shared with other demands (lower availability in case of multiple failures but lower redundancy)~\cite{Shen2014,Liu2013}. Another technique to offer protection in mesh topologies aiming to reduce the spare resources are p-cycles, which consists of finding cycles offering protection in case of link failures (one of those cycles can cope with any on-cycle and straddling link failures). p-cycles offer ring like recovery speed and mesh-like capacity efficiency and they have been shown to reduce significantly the resource redundancy can also be easily applied to optical networks \cite{Schupke2002, Ji2013}.

In contrast to protection that has to be computed before the failure event, restoration is triggered once the failure(s) has/have been identified. Based on the available resources, the interrupted demands are aimed to be restored~\cite{Liu2015}. Although restoration requires lower resource redundancy, it has longer interruption time as it can be triggered only once the failure(s) has/have been identified. Some works have addressed these schemes not only at the optical layer but at higher layers (e.g., IP)~\cite{Papanikolaou2017, Chiu2011}.

\subsubsection{Placement Problems}
This section briefly presents some placement problems that optical network planners have to address.
\begin{itemize}
    \item Wavelength Converter placement problem: The lightpaths configured in transparent optical networks have the same wavelength/spectrum along the path. This wavelength/spectrum continuity constraint increases the blocking probability of the demands. The problem could be solved by installing converters at all nodes, but this significantly increases the costs. Therefore, in order to find a compromise between the number of converters and reducing the blocking probability, the wavelength converted placement problem has been addressed~\cite{WC1, WC2, WC3, WC4}. The proposed solutions differ depending on the traffic scenario (static vs. dynamic demands) and on the methodology (tabu-search, analytical, machine learning, Integer Linear Programming (ILP), etc.). 
    
    The objective of the wavelength converter placement problem is to minimize the number of converters in the network given the network topology, the lightpath demands, and the number of ports supported by the wavelength converters. The problem finds the paths for the requested demands and the wavelength(s) assigned to them, assuring the wavelength continuity constraint is guaranteed unless there is a wavelength converter in the node with enough capacity(ports) and within the conversion range~\cite{WC3}. 
    \item Regenerator placement problem: Optical regenerators allow overcoming the maximum optical reach limitation. Optical regenerators are classified into three categories: (i) 1R regenerators are able to re-amplify the optical signal, (ii) 2R regenerators are able to re-amplify and reshape the optical signal, and (iii) 3R regenerators are able to re-amplify, reshape and re-time the optical signal~\cite{Boscolo2005}. Optical regenerators rely on non-linear effects such as self-phase modulation, cross-phase modulation, and Four Wave Mixing. However, thanks to the cost decrease and the flexibility of optical transponders, B2B regenerators~\cite{Simmons2008} are most used as they also perform all the regenerator functions as well as wavelength conversion. These capabilities significantly increase the network flexibility as longer lightpaths with high bitrates can be set, which allow reducing the number of required transponders, specially in core networks with a short diameter (e.g., Germany50, which has a diameter of 934km)~\cite{Khomchenko2021,Saquib2022}. The regeneration placement problem identifies the network nodes that should host regenerators such that all the demands can be set. There is a second problem referred as regenerator assignment problem, which deals with the assignment of regenerators to the demands such that the blocking probability is zero~\cite{Cavalcante2019, deSilva2020}. In that case, the problem results in the dimensioning of the regenerators at each regeneration location and therefore, it is directly related to the network investments.   
    
\end{itemize}

\subsection{Optical Network Control and Management}

Optical network operators provision their services, i.e., lightpahts, through their control and management planes. The previous section has summarized different approaches to compute the resources that have to be reserved for a given demand. However, these resources have to be mapped to the transceiver's configuration and to forwarding rules at the intermediate ROADMs along the path. The management plane deals with the Fault Management, Configuration, Accounting, Performance and Security (FCAPS) tasks~\cite{ISO}. The Simple Network Management Protocol (SNMP) deals with the data collection, data organization, and monitoring of the different network devices. On the other hand, the control plane deals with the functions associated with the lightpath(s) provisioning. The control plane needs standard interfaces to operate across domains and 
ensures vendor inter-operability~\cite{Casellas2018}. Hence, the functions of the control plane are: element addressing; dynamic resource discovery; automatic topology and reachability; RSCA computation and service provisioning with recovery (protection and restoration) schemes while ensuring efficient resource usage~\cite{Casellas2018}. The control plane can be centralized or distributed. The former (e.g., pure Software Defined Networking (SDN)) has a controller hosting all the functions and interacting with each of the network components. The later (e.g., Generalised Multi-Protocol Label Switching (GMPLS)) places a controller at each node so that a communication between all the controllers is required in order to have a consistent network operation. Both alternatives have advantages and disadvantages. A centralized control plane merges all the functions in a single entity, which requires less state synchronization but it presents a critical failure location. Furthermore, some functions (e.g., dynamic restoration) perform better in distributed control planes, as the state information is disseminated along the network. However, more synchronization is required. Hence, real deployments implement a hybrid control plane.  

Furthermore, network virtualization applied to optical networks allows isolating different slices while coping with the network resources in a flexible manner~\cite{virt}.
In other words, network virtualization provides flexible network resource provisioning such that optical network operators can allocate the optical resources to different virtual optical networks, i.e., clients. Each client can have the control of the assigned resources but without being able to disturb other virtual networks, even if they are sharing the same physical network.
Abstraction is a key concept for optical network resource provisioning, which allows hiding the physical layer information from the resources that should be provisioned. One abstraction technique is participation, where the resources of e.g., a switch or a fiber are sliced into multiple parts, so that each virtual network can use a different part. The opposite technique is aggregation, where several resource parts are merged into one bigger virtual entity.
\begin{figure}[h]
\centering
\includegraphics[width=0.7\textwidth]{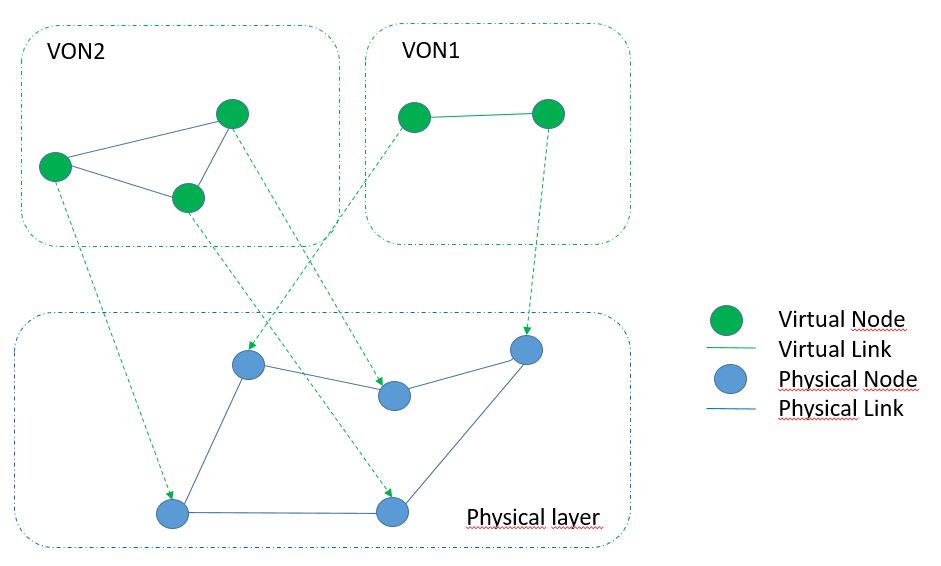}
\caption{Example of two virtual optical networks (VON) on the same physical layer}
\label{fig:RWA2}
\end{figure}
Last but not least, NFV allows replacing network functions implemented in custom hardware appliances with softwarized functions that can be run on different virtual machines at different servers. Some examples of NFV applied to optical networks are: the global transport SDN proposed by the Optical Internetworking Forum (OIF) and the Open Networking Foundation (ONF), the control of transport networks framework proposed by Internet Engineering Task Force (IETF), or the Path Computation Element as a Network Service architecture proposed by IETF.

\section{Optical Data Center Interconnects}
Driven by growing importance and popularity of online services executed at data centers (DCs) the related traffic demand is drastically increasing. 

Consider for example a DC with 1,024 racks, with 20 servers in each rack, and with 16 processor cores, 64 GB of memory, and 48 TB of storage in each server. This corresponds to a DC infrastructure having 327,680 cores, 1.31 PB of memory, and 983 PB of storage. Such DC infrastructures are commonly owned and operated by companies such as Google, Microsoft, Facebook, etc., to provide modern web and mobile applications such as social networking, search, maps, and video streaming, to give a few examples. Besides, most of the services handled by DCs come with very strict latency, reliability, and security requirements. This brings serious challenges for interconnection between servers and racks inside the DCs.

The main challenge is to provide sufficient capacity that can support the growing intra-DC traffic volumes. Upgrading existing infrastructures with additional electronic equipment will lead to a drastic increase of energy consumption and latency, as well as lower reliability performance.

Large capacity demand in DC inteconnects leads to the need for deploying optical networks inside the DCs. Only optical communication technology can offer a sufficient capacity. However, traditional DC network architectures limit the use of optics to the physical cables only, while optical interconnects offer much higher transmission and switching capacity, as well as much lower energy consumption per bit compared to the interconnects based on the electronic packet switching.  

In addition, the increasing service demand leads to a need for upgrading DCs with more computing resources. Therefore, the efficient utilization of computing resources in terms of central processing units (CPUs), memory and storage units gains high importance. Currently, an inefficient utilization of compute resources comes from the deployment of integrated servers, where the number of CPUs, memory and storage units is pre-defined. In a big DC thousands of blade servers are interconnected in the network. A blade server contains fixed amount of different resources (i.e., CPU, memory, storage) integrated together on the server’s bus. However, the services running on the servers are diverse and may require different ratios of these resources.

The mismatch between the diversity of resource required and the fixed amount of resources integrated in the physical blade servers may lead to so called resource stranding\cite{DC1}. Resource stranding means that the  applications running in a server have used up one type of resource while the other types of resources are still idle and cannot be used. Therefore, it may be beneficial to disaggregate different resources in DCs and utilize them according to the demand. However, it comes with tough requirements on the capacity and latency of the network interconnecting different types of computing resources, which can only be satisfied by photonic technologies.

To illustrate the current developments and the possible ways to address the aforementioned challenges, this sub-chapter presents various optical DC network architectures along with the future developments in terms of possible solutions for energy saving, increased capacity, and resource disaggregation. 

\subsection{Optical DC network architectures}
\begin{figure}[h]
\centering
\includegraphics[width=0.6\textwidth]{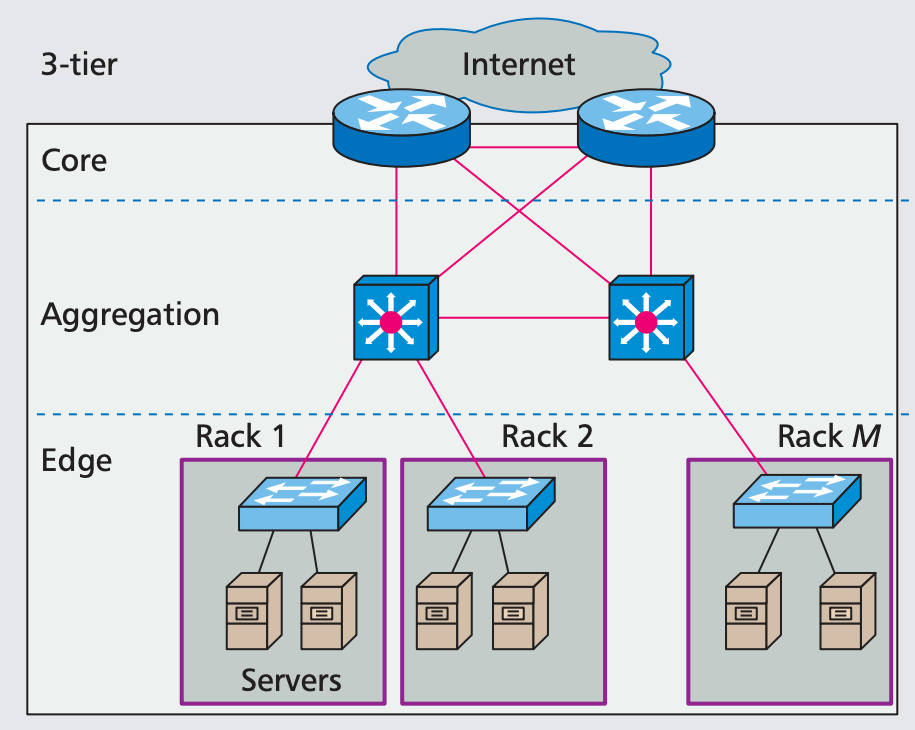}
\caption{Hierarchical DC network architecture~\cite{DC2}}
\label{fig:DC1}
\end{figure}

Figure\ref{fig:DC1} illustrates a hierarchical DC network architecture consisting of three levels, referred to as tiers, i.e., access (edge) tier, aggregation tier, and core tier.

The servers in the DCs are grouped and mounted in racks. The access tier is responsible for interconnecting the servers within the rack. Therefore, the access (edge) tier is often referred to as top of the rack (ToR) switch. The ToR switches are further connected to the higher tier. In the current DC networks, the optics is mainly used for transmission, while switching is performed by electronic packet switches (EPSs). One of the major issues of such EPS-based DC network is the high power consumption. Therefore, the next step in the evolution of the DC network architecture is reducing the number of EPSs by merging aggregation and core tiers and by replacing electronic switching with optical switching. 

Several architectures utilizing optical switching technologies have been proposed for the aggregation and core tiers. 

\begin{figure}[h]
\centering
\includegraphics[width=0.6\textwidth]{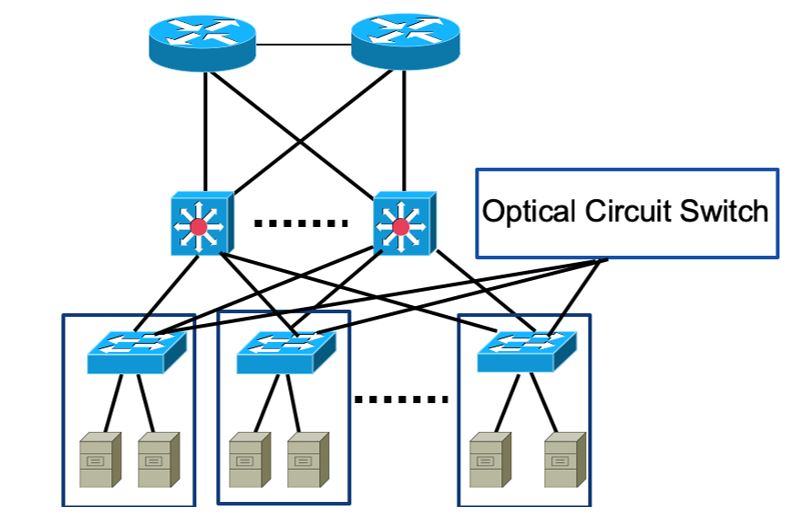}
\caption{Helios, A Hybrid Electrical/Optical Switch Architecture for DCs ~\cite{DC3}}
\label{fig:DC2}
\end{figure}

Figure~\ref{fig:DC2} shows an example of a hybrid solution~\cite{DC3}, where electronic packet switching is applied for small-volume data flows (referred to as mice flows) and optical circuit switching technology is applied for high-volume data flows (referred to as elephant flows). In addition, the aggregation and core tiers are merged in the core switches, which can also reduce the energy consumption of the switching system.

\begin{figure}[h]
\centering
\includegraphics[width=0.6\textwidth]{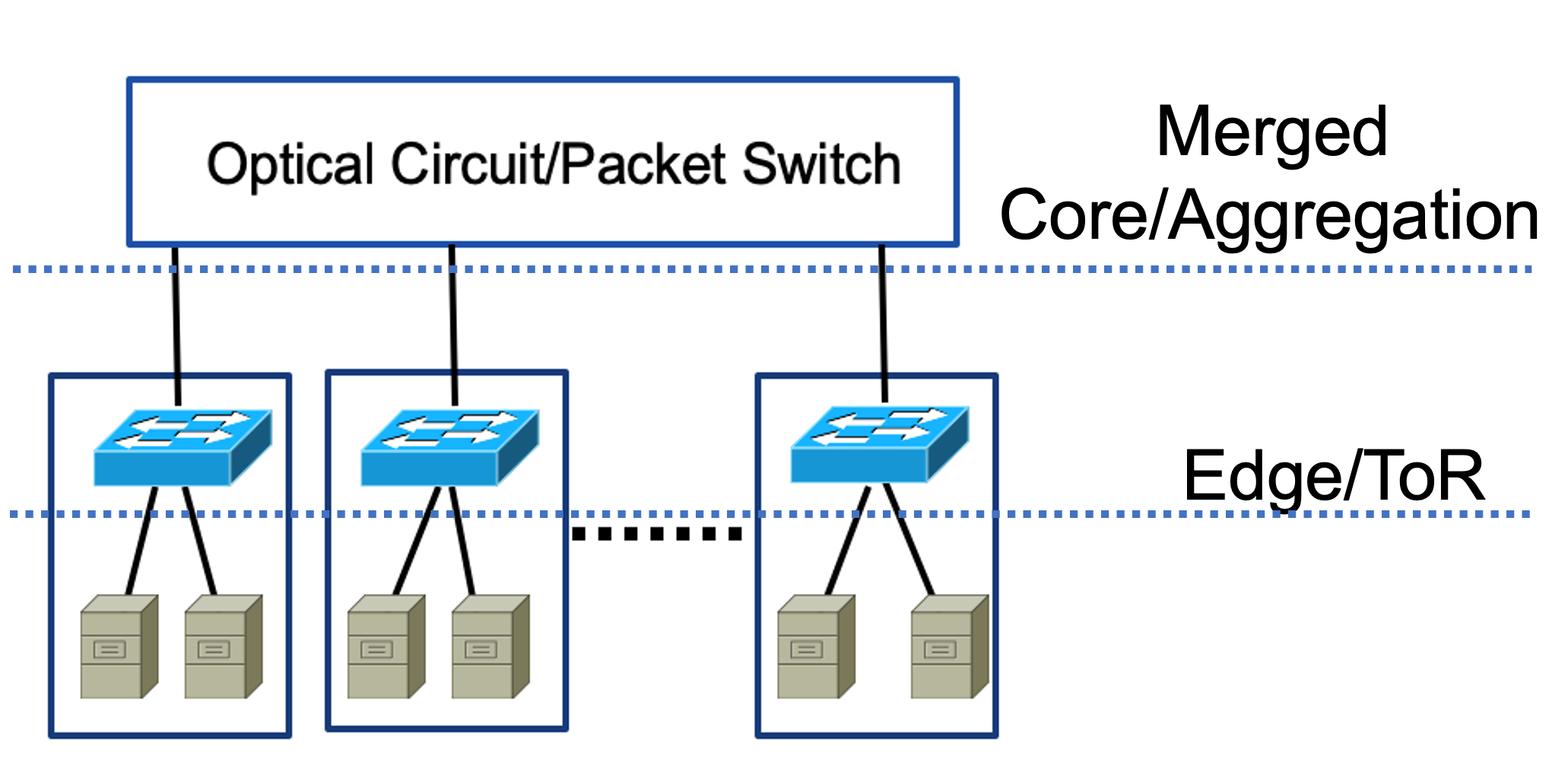}
\caption{DC network architecture with merged core/aggregation tier based on optical switching~\cite{DC4}}
\label{fig:DC3}
\end{figure}

In order to further reduce energy consumption and cabling complexity merging aggregation and core tiers by all-optical switching architectures has been proposed, as shown in Figure~\ref{fig:DC3}~\cite{DC4}. The optical switching can be based on either optical circuit switching~\cite{DC5} or optical packet switching~\cite{DC6}. In both cases the optical switching architectures require active optical switching devices, e.g., wavelength
selective switches (WSSs) and optical space switching matrices. Additionally, in optical packet switching the buffering for contention resolution is performed in electronic domain.

\begin{figure}[h]
\centering
\includegraphics[width=0.7\textwidth]{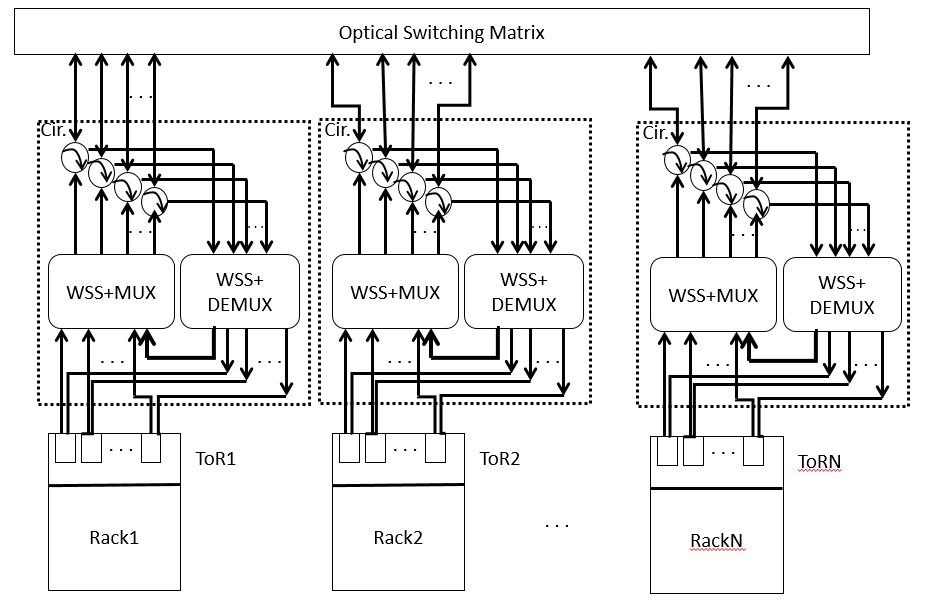}
\caption{Optical Switching Architecture (OSA), the optical circuit switching based DC network~\cite{DC5}}
\label{fig:DC4}
\end{figure}

\begin{figure}[h]
\centering
\includegraphics[width=0.5\textwidth]{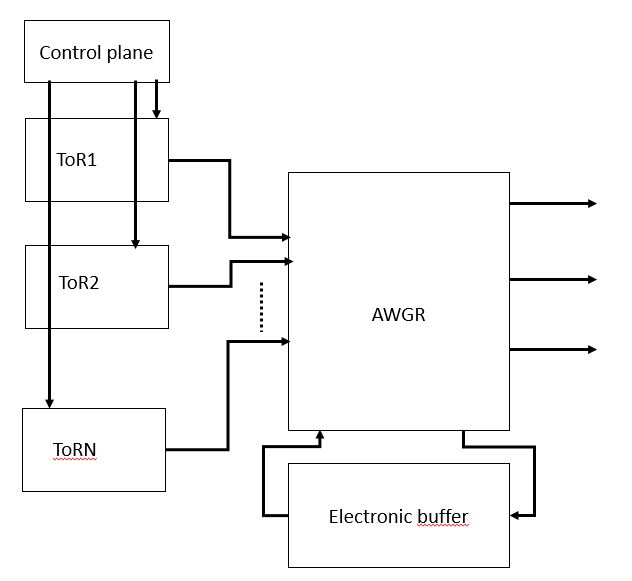}
\caption{Low-Latency Optical Switch for DCs based on AWGR (LIONS)~\cite{DC6}. AWGR: arrayed waveguide grating router.}
\label{fig:DC5}
\end{figure}

An example of the circuit switching based optical DC network architecture~\cite{DC5} referred to as Optical Switching Architecture (OSA) is shown in Figure~\ref{fig:DC4}, while Figure~\ref{fig:DC5} shows an example of an optical packet switched DC network~\cite{DC6}  referred to as low-latency interconnect optical network switch (LIONS), where packet contention resolution is performed by electronic buffers, and therefore extra optical/electrical (OE) and electrical/optical (EO) conversions are needed.

It has been shown that applying optical switching technologies in core/aggregation tier can achieve almost 50\% reduction of the energy consumption per bit compared to DC networks based solely on EPSs~\cite{DC8}. The reason for this limitation is that a lot of energy is consumed by the EPSs in the edge, i.e., at ToR. Therefore, to further reduce the energy consumption, passive optical interconnects (POIs) at ToR have been proposed in~\cite{DC7}. It has been shown that these architectures are able to reduce energy consumption per bit by a factor of 10~\cite{DC7}. 
The POI concept has been initially introduced to replace EPSs at ToR of the large DCs as shown in~\cite{DC7}. However, the research efforts published in \cite{DC8} addressed the extension supporting medium-size DCs and/or the aggregation/core tier of the large DCs. The main idea of the POI is to realize interconnects by using only passive optical components, such as optical power splitters/couplers and wavelength splitters, i.e., arrayed waveguide gratings (AWGs). 

\begin{figure}[h]
\centering
\includegraphics[width=0.99\textwidth]{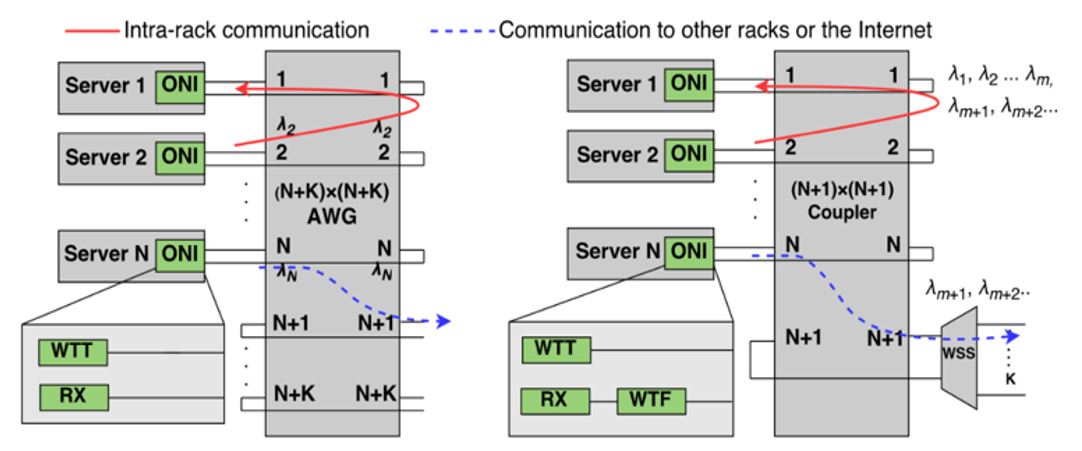}
\caption{AWG based POI (left) and coupler based POI (right). WTT: wavelength
tunable transmitter, AWG: arrayed waveguide grating, ONI: optical network interface, RX: receiver, WTF: wavelength tunable filter, WSS: wavelength selective
switch.~\cite{DC7}}
\label{fig:DC6}
\end{figure}

Figure~\ref{fig:DC6} shows two architectures of POI on ToR~\cite{DC7}, namely AWG based (on the left) and coupler-based POI (on the right). The first option requires N+K wavelengths to support up to N servers within a rack, where N wavelength channels are assigned for communication between any pair of servers within the rack and K wavelengths are reserved for communication between the servers and uplink ports. A strict wavelength plan is needed following the cyclic property of the AWG. It reduces the architecture flexibility. In the second option, an (N+1)x(N+1) coupler also interconnects N servers in a rack. Each optical network interface (ONI) on the server is connected to a pair of the input and output ports of the coupler. The last input/output port of the coupler is reserved for connection to a wavelength selective switch (WSS) that chooses the wavelengths assigned to the communications between the servers and uplink ports and blocks the wavelengths for the communications between any pair of servers within the rack. The WSS here is an active component, so the overall interconnect architecture is not fully passive. To keep the architecture strictly passive, the WSS can be eliminated. However, it would lead to inefficient wavelength utilization. The broadcast nature of the coupler-based POI gives this architecture higher flexibility in wavelength allocation compared to the AWG based POI where a fixed wavelength plan is needed. On the other hand, an additional component, e.g., a wavelength tunable filter (WTF), is required in ONI in order to select the specific wavelength assigned to the certain communication. 

\begin{figure}[h]
\centering
\includegraphics[width=0.45\textwidth]{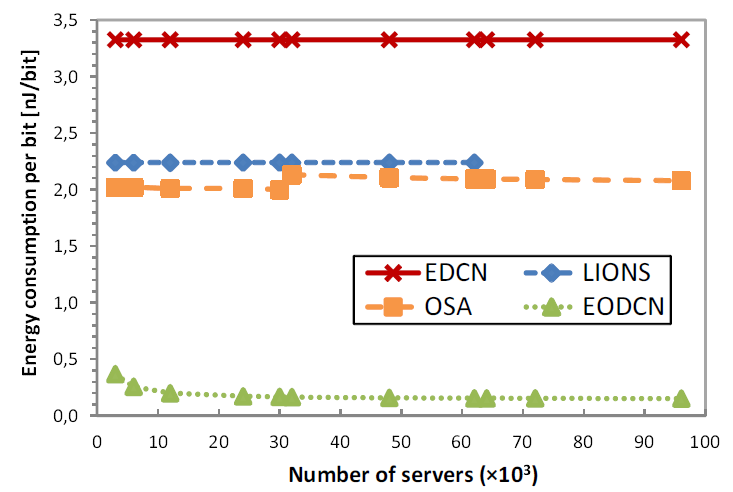}
\includegraphics[width=0.45\textwidth]{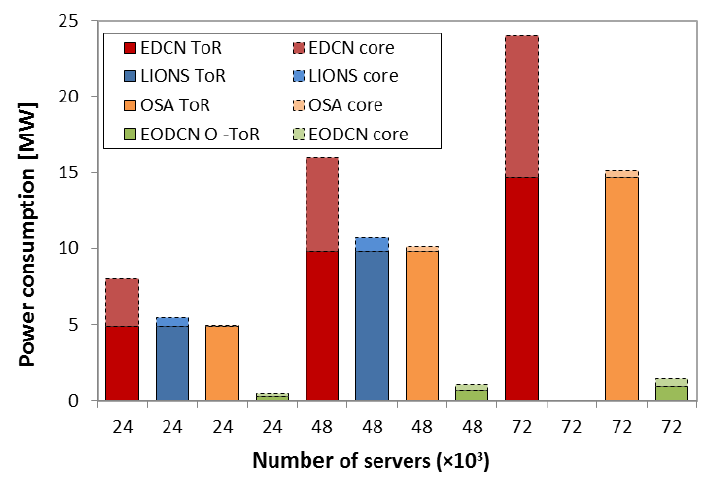}
\caption{Comparison of energy consumption per bit and power consumption as a function of number of servers in DC networks based on different switching technologies.
EDCN: DC network based on EPSs 
EODCN: DC network based on flex grid in the core tier and POI at ToR
.~\cite{DC8}}
\label{fig:DC7}
\end{figure}

The POI architectures illustrated in Figure~\ref{fig:DC6} have been shown to be very energy efficient. 
This was confirmed by the results presented in Figure~\ref{fig:DC7}~\cite{DC8} showing the comparison in terms of energy consumption per bit (on the left) and power consumption (on the right) between OSA and LION architectures (incl. EPS based ToR) with the DC network architecture based on flex grid optical switch in the core tier and POI at ToR, referred to as EODCN. A DC network based on EPSs is used as a benchmark.

\subsection{Future Developments}
As discussed previously, to address the drastically increasing intra-DC traffic in a cost and energy efficient way, further improvement of scalability, resource utilization and reduction of cabling complexity is needed. This sub-chapter highlights resource disaggregation and spatial division multiplexing (SDM) as powerful means for solving resource utilization and capacity problems in future DC networks.
\subsubsection{Resource disaggregation}

As already mentioned, the major reason that limits resource utilization in DCs is a possible mismatch in terms of compute resources required by the various applications and the amount of CPU, memory and storage integrated in the physical blade servers. This problem can be mitigated by disaggregating different resources in DCs and utilizing them according to the demand. 
In contrast to the integrated blade server, where fixed amount of compute resources is integrated in the server chassis, the disaggregated DCs contain decoupled resources held on different resource blades. The resource disaggregation has a great potential to improve resource utilization, but it requires ultra-high bandwidth and ultra-low latency links interconnecting different types of resources, bringing a significant challenge for intra-DC interconnects.

\begin{figure}[h]
\centering
\includegraphics[width=0.99\textwidth]{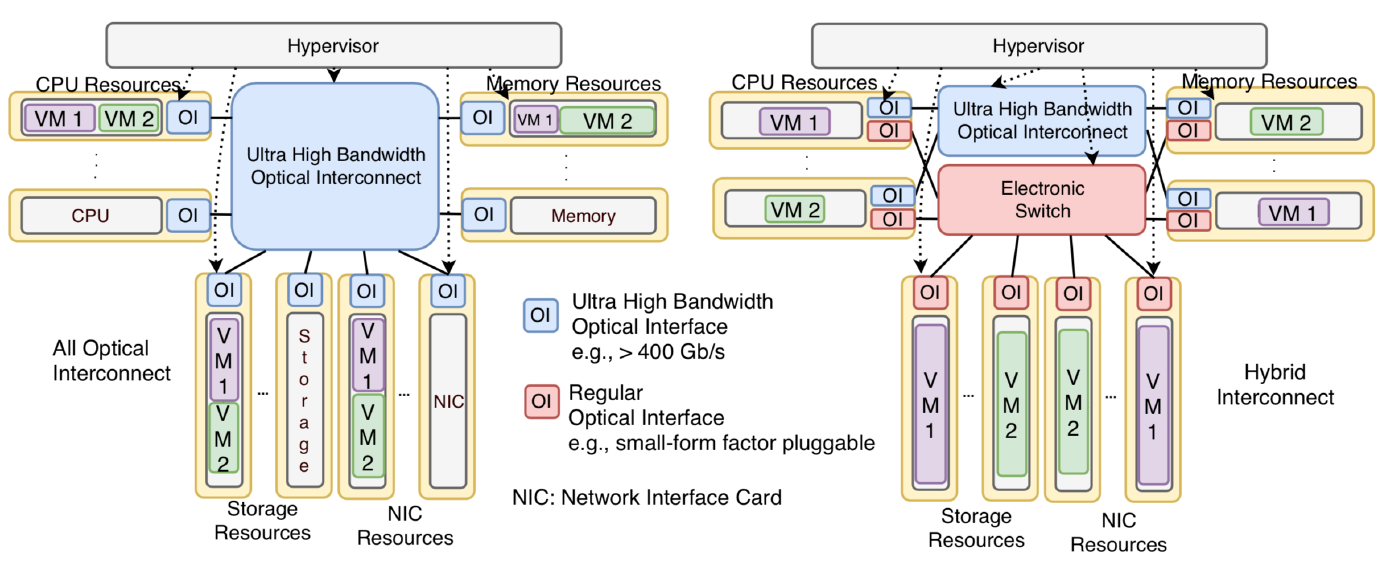}
\caption{Fully disaggregated architectures with all-optical (left) and hybrid (right) interconnect approaches
.~\cite{DC9}}
\label{fig:DC8}
\end{figure}

Recently, partial resource disaggregation was widely deployed in DCs, in which the storage is decoupled and connected to the other computing resources (such as CPU and memory) through external switch fabrics. A more aggressive solution, i.e., fully disaggregated architecture (see Figure~\ref{fig:DC8}), in which every type of resources, including CPU, memory, storage, NIC, and accelerators, such as graphics processing units (GPUs) (not shown in the figure), are completely decoupled from each other~\cite{DC9}. It enables DC operators to replace/upgrade any type of resources separately and when necessary, leading to significantly improved resource utilization and availability performance. In this scenario a hypervisor~\cite{hyp} can be used to monitor and manage the virtual machines (VMs) running in different resource blades. The CPU-related interconnects, i.e., the ones between the CPU and memory, as well as between the CPUs are most bandwidth-demanding in this kind of architectures. These connections usually require peak bandwidths higher than 400 Gb/s and therefore need optical interconnects offering ultra-high capacity. The main difference between the two approaches shown in Figure~\ref{fig:DC8}, namely the all-optical and hybrid, is a need for the electronic switch in the hybrid architecture as well as the required additional regular optical interfaces (OIs). This difference may cause a substantial increase in energy consumption and extra complexity on cabling. On the other hand, the communication coordination in the all-optical architecture may be more complex, because every communication link from/to a resource blade is handled by the single ultra-high capacity optical interface (OI). This drawback may lead to a negative impact on the latency performance. The memory blades also need special considerations, so that the ultra-high bandwidth CPU-memory communication does not use up the OI bandwidth all the time, starving the memory-storage and memory-NIC communications. Such trade-offs need to be fully explored to determine the proper architecture for further investigation. 

Moreover, it is important to develop resource allocation algorithms for VM deployment by considering the low-latency requirements to jointly optimize energy consumption and usage of computing, storage and communication resources in the time, spectrum and space domains. Meanwhile, managing the traffic burstiness in DCs is extremely challenging. Recent advances in machine learning, such as supervised and reinforcement learning strategies, might be helpful to might be helpful to construct a traffic arrival prediction system that is capable of learning from past traffic flows. Accurate prediction of the DC traffic is critical, not only to efficiently manage, provision, and reconfigure network resources and services, but also to mitigate the optical system capacity limitations.

\subsubsection{Spatial division multiplexing (SDM)}
\begin{figure}[h]
\centering
\includegraphics[width=0.8\textwidth]{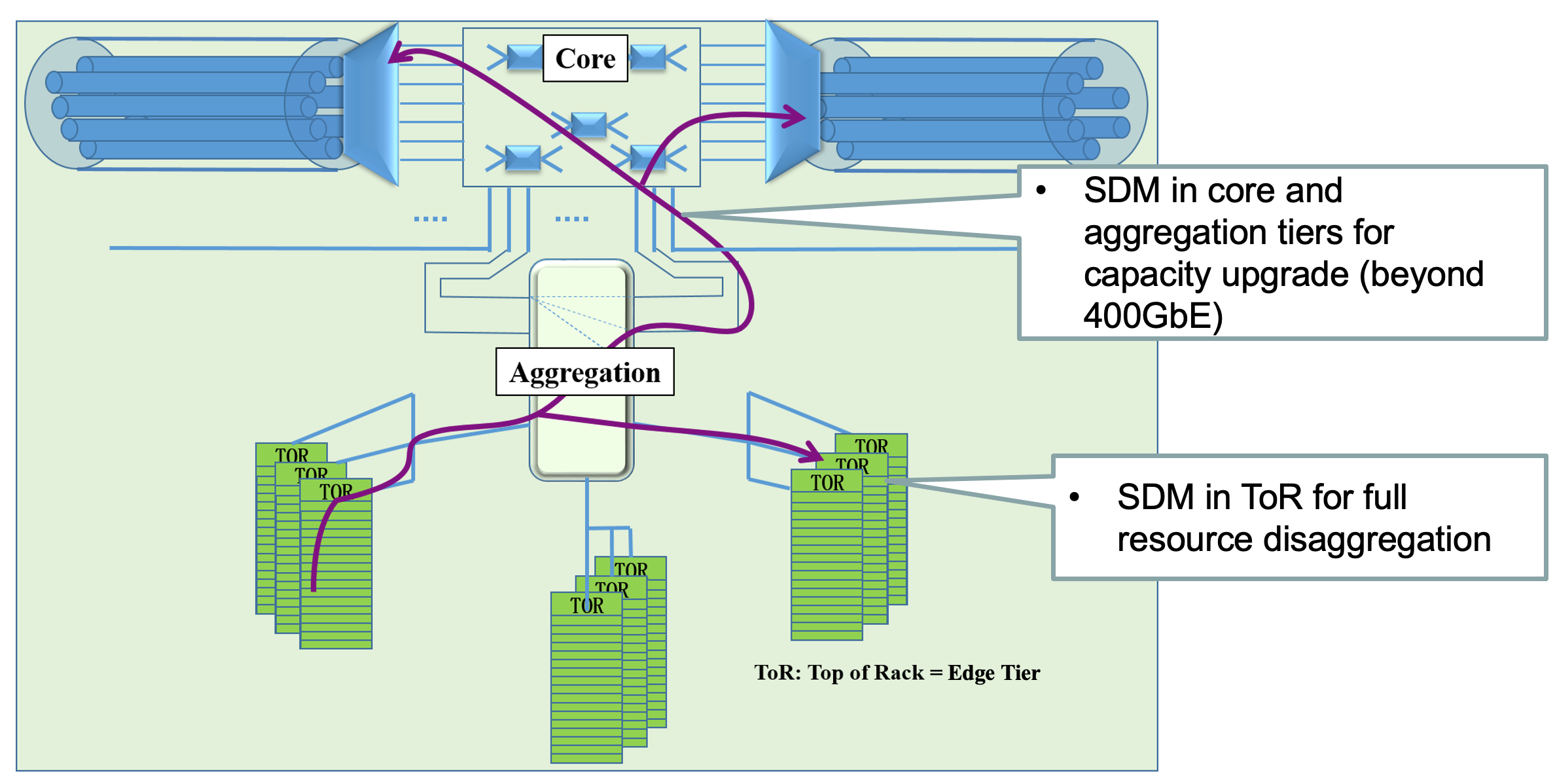}
\caption{Potential placement of SDM techniques in future DC networks 
.~\cite{DC10}}
\label{fig:DC9}
\end{figure}

The current intra-DC networks use either individual fibers or fiber ribbons, which are costly, bulky, hard to manage and not scalable. SDM fibers, in particular multi-core fibers (MCFs), are regarded as a feasible and efficient way to address these problems. Moreover, as previously mentioned, ultra-high capacity is required to support fully disaggregated DCs, which is an additional driving force for  implementing SDM fibers in DC networks. Figure~\ref{fig:DC9} shows the potential placement of SDM techniques in future DC networks. \cite{DC10} provides a detailed overview of the components and transmission options outlining the SDM development trend. This sub-chapter focuses on the technical directions related to the co-existence of SDM and wavelength/spectrum division multiplexing in attempt to address ever-increasing capacity demand and resource utilization problem in DC networks.

\begin{figure}[h]
\centering
\includegraphics[width=0.99\textwidth]{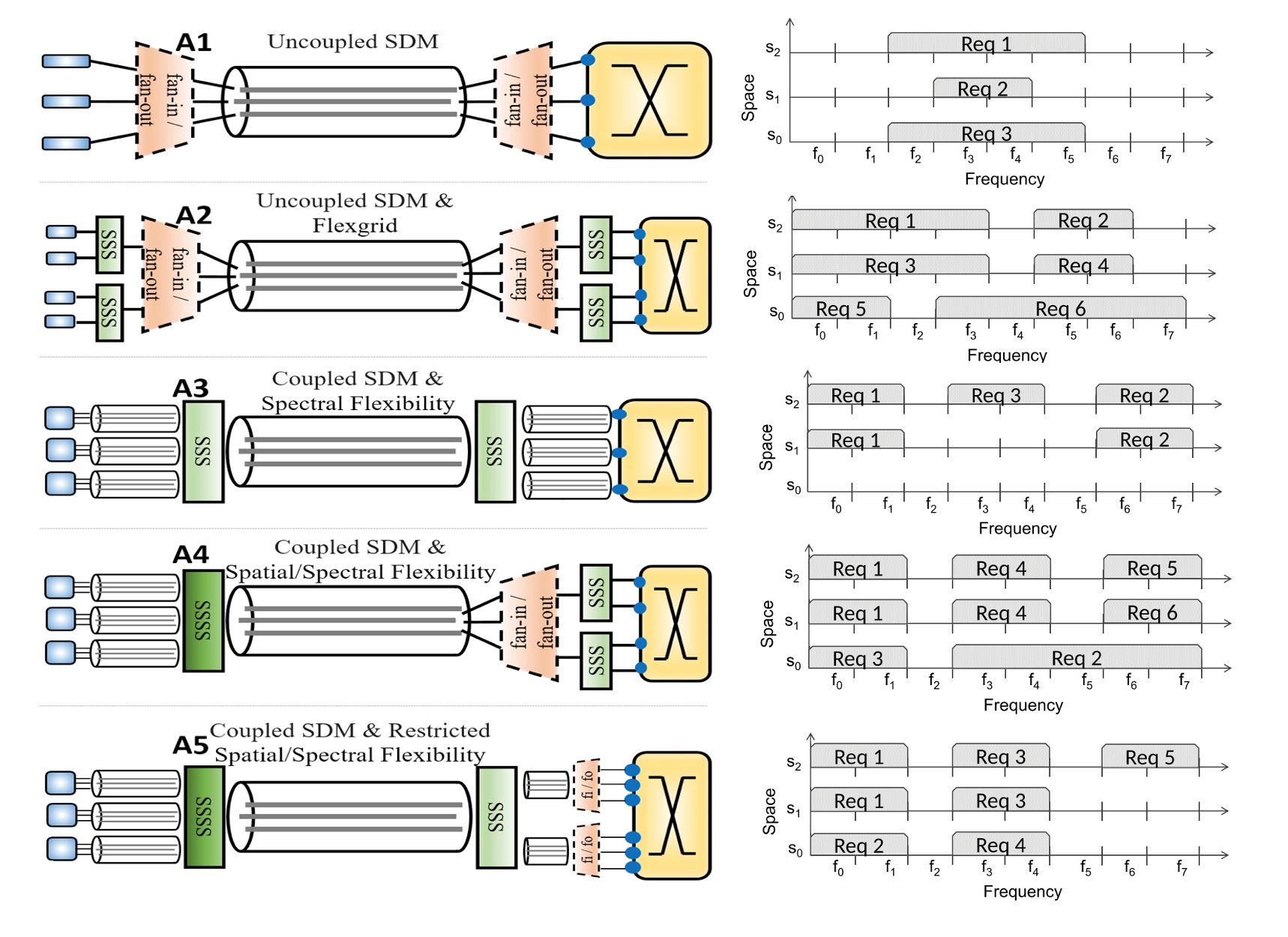}
\caption{Combined SDM and spectrum multiplexing based optical interconnect options (from top to bottom): A1 uncoupled SDM, A2 uncoupled SDM and flex-grid, A3 coupled SDM with spectral flexibility, A4 coupled SDM with spectral and spatial flexibility and A5 coupled SDM with restricted spectral and spatial flexibility. SSS: spectrum selective switch, SSSS: spectrum and space selective switch, fi/fo: fan-in/fan-out, Req.: connection request.~\cite{DC10}}
\label{fig:DC10}
\end{figure}

Figure~\ref{fig:DC10}  shows 5 candidate SDM-based intra-DC network architectural options (more details about this figure combined SDM and spectrum multiplexing based optical interconnect options can be found in ~\cite{DC10, DCF}):

\begin{itemize}
    \item \textbf{Architecture A1}  referred to as the uncoupled SDM architecture, applies only switching between independent spatial elements. 
    \item \textbf{Architecture A2}  referred to as the uncoupled SDM with flex-grid architecture, allows each spatial element to work as a separate flex-grid transmission medium.  As A2 keeps each spatial element independent, it requires spectrum selective switch (SSS) for each spatial element and two spatial fan-in/fan-out (fi/fo).    
    \item \textbf{Architecture A3}   referred to as the coupled SDM with spectral flexibility architecture, expands spectral superchannels in the spatial domain to create spectral-spatial superchannels with spatial coupling. Therefore, two large SSSs need to be employed.    
    \item \textbf{Architecture A4}  referred to as the coupled SDM with spectral and spatial flexibility architecture, offers an unrestricted flexibility in both spectral and spatial domains, which leads to the highest flexibility degree compared to the other architectures. On the other hand, A4 requires a large spectrum and space selective switch (SSSS) to interconnect the SDM fiber. Moreover, because of the high flexibility, advanced resource allocation approaches are needed to ensure efficient resource utilization.
    \item \textbf{Architecture A5}  referred to as the coupled SDM with restricted spectral and spatial flexibility architecture, can establish flexible spectral-spatial superchannels but needs to make sure that the spectral-spatial superchannels belonging to the same spectral group utilize the same spectrum resources at all spatial elements. A large SSSS is required to connect the SDM fiber. Compared to the unrestricted flexibility in A4, the restricted flexibility in A5 reduces the complexity of the resource allocation and cost of the hardware, reflected by the lower amount of SSSs. 
\end{itemize}

Performance comparison of the combined SDM and wavelength/spectrum division multiplexing architectures shows that higher resource allocation flexibility leads to more efficient resource utilization. However, it should be noted that the highest flexibility comes with the highest cost, which is not always necessary, particularly in relatively small DCs. Meanwhile, SDM might introduce extra latency, not only because of extra complexity caused by the signal processing process for handling inter-spatial channel impairments, but also by potential additional level of switching in the space domain (e.g., shown in A4 and A5). This latency impact should be further investigated. 
\section{Conclusion}
This chapter highlighted the importance of optical networks for the further development of digitalization, crucial for the societal advance.

First, the chapter provided a brief overview of technology and architectural evolution of optical access networks, driven initially by increase in broadband traffic requirements and more recently by requirements for 5G and beyond services. While the technology has evolved steadily towards higher data rate per channel and multiple wavelengths, the proposed architecture has changed in different ways. For example, the concepts of ultra dense WDM, Long-Reach PON and mesh PON architectures have been described. The evolution towards virtualized implementation has also been presented. It can bring new, unprecedented flexibility to access networks, an important feature to support evolutions towards novel, often unpredictable, services and applications.

In optical core networks, the fast development of optical network devices makes it possible to both increase transmission capacity and improve optical network flexibility, resource efficiency and performance. Different planning problems have been discussed with the emphasis on the optimal resource allocation and reliability performance. Moreover, different options for optical network control and management have been briefly described.

Finally, the main challenges for intra-DC interconnects, such as increasing traffic demand and power consumption, have been highlighted. These challenges lead to a necessity of applying optical interconnects inside the DCs. Some ideas for optical intra-DC network architectures have been reviewed and compared. and resource disaggregation and SDM have been identified as future developments to address the rapidly growing traffic demand and to improve power and cost efficiency.       

%
%
%
%

\end{document}